\documentclass{ws-p8-50x6-00}

\begin{document}

\title{Probing colour separate singlet states in $e^+ e^-$ annihilation at
$Z^0$ pole}

\author{Li Shi-yuan, Shao Feng-lan and  Xie Qu-bing}

\address{Department  of Physics, Shandong University,\\
 Jinan, Shandong, P. R. C., 250100\\ 
E-mail:tpc@sdu.edu.cn}


\maketitle

\abstracts{By selecting two-jet like events and employing sensitive
observable, 
we find that two kinds of colour connections at the hard-soft interface lead
to significantly differences in  hadron states. 
}

\section{Introduction}
In high energy strong interaction processes, e.g., $e^+e^- \to
hadrons$,  the confinement mechanism demands that at 
 the end of the PQCD cascade the
 partons should be connected, 
phenomenologically, as the 
strings in Lund model, or  clusters in Webber model\cite{lund}$^,$ \cite{webber}
from which 
the hadronization starts. This `topology' of the parton system on  the
hard  and soft interface   is beyond PQCD
approach\cite{wangqunprd}'\cite{wangq}.

In the `standard models' of parton cascade and 
hadronization\cite{lund}$^,$ \cite{webber},  
the $N_C \to \infty$ approximation($N_C$  is the number of colour) 
is applied.
This approximation leads to the result that 
at the end of the PQCD cascade, there is a direct correspondence
between the parton states and the string/cluster chain states.
Such a simple but phenomenological
approximation conceals the complexity of the colour connections on the
hard and soft interface. It
had  not yet met any real challenge mainly because the `standard
models'\cite{lund}$^,$ \cite{webber}
which employ the large $N_C$ approximation can reproduce most of the
 experimental data especially  the global properties 
  until now.

On the other hand, the straightforward `group calculus'
(i.e., direct product and
reduction of the representations of the $SU_C(3)$ group) shows that  
the   colour
structure of a multi-parton state is copious and complex as long as 
 $N_C=3$(see, e.g.,  \cite{wangqunold}$^,$ \cite{wangqunprd}'\cite{wangq}).
 This implies that  the colour connections of the  multi-parton system on
the hard and soft interface 
can be of many  kinds.
The  traditional one, which is employed in the `standard models' and will be  
called 
colour singlet chain states  hereafter, is only one kind among them. 
The PQCD calculations show that, projected     
onto the colour space of PQCD final state,  such  kind of colour connection 
can never have the 
probability 1 as long as $N_C=3$  and that when parton number becomes larger the
probability is reduced\cite{jinyi}.
Another kind of colour connections is colour separate singlet states, with the
distinguish feature that several gluons form  colour singlet 
clusters and can hadronize independently. 
 Since only 3 kinds of colour(8
kinds of `double colours'), when parton number is large\footnote{as 
calculated by JETSET, the average numbers of
gluons are: $\sqrt{s}=91 GeV, ~<N> \sim 6;~ \sqrt{s}=200 GeV, ~<N> \sim
9; ~ \sqrt{s}=1 TeV, ~<N> \sim 17$}, 
it is inevitably that more and more gluons have
identical colours, 
and the colour separate states could  have large
probability.

We should  notice that, from our previous works\cite{wangqunold}' \cite{wangq}' \cite{wangqunprd}, 
the above 
two kinds of colour configurations, 
colour singlet chains with various   orders for the gluons in the chain, 
 colour separate singlets
with different gluons combined into singlet clusters, are not orthogonal to
each other as long as $N_C=3$ and within PQCD framework. 
 They belong to different complete sets of bases for the
PQCD final state colour space. Hence  PQCD sees no differences among them.
On the other hand,  different colour connections mean different string/cluster
  states.  
This implies that different colour connections  can lead to differences 
in  handron states.
Furthermore,  the
cross section  of any colour connection we calculated in PQCD by projecting  
it onto  
the PQCD colour space
 could have no realistic meanings. Their relative probabilities
 should depend on
the unknown NPQCD mechanism.
To analysize these issues, one needs:  

1) To confirm  if  the colour singlet chain states and the colour separate
singlet states really do not equal to each other by seeing if they lead to
the differences in  hadron states for the same parton state and hadronization
scheme.

2)If they do not equal, to see how does  nature chooses among them (maybe) by
the(unknown) NPQCD laws.

Because of the lack of the knowledge of NPQCD, 
we can not answer these questions
from the QCD Lagrangian, but  employing  phenomenological  models.
The straightforward way is to modify the 
event generators(e.g., JETSET as discussed following):  
 Only  the colour connections changed from colour singlet chains to colour separate singlets 
while the other parts (parton
cascade and hadronization scheme, etc)  kept. Then  one  can use the 
modified  generators to produce  the
hadron states to see if there are some deviations  from the corresponding traditional results. 
Further step is to compare with the data especially those  the 
unmodified generators  can
not explain, to extract   the relative weights of all the connections.

There have been  some works along this route(e.g.,  \cite{gus96} and refs. therein). 
Recently we constructed a  colour separate singlet model\cite{wangq}by
modifying the standard JETSET7.4 code. Hereafter, we use CS(Colour Separate)
 for the modified  JETSET,
and CC(Colour singlet Chain) for the standard.  
However,   the results of the CS  seems an ambiguity: 
Though the colour connection is different from that of CC,  
the  observables   describing the hadron states 
 averaged among  unbiased  events
have no significant 
differences from  those of th CC 
provided that a T-measure to weight the colour separate  states
(Hence `short strings' are
favoured(see, e.g., \cite{gus96} and refs. therein). 
Considering the uncertainties in the model, 
 somehow we can say that for unbiased  events, these two kinds of colour
connections can be thought of as leading to the same hadron states\cite{wangqunprd}'  \cite{shaofenglan}.

The  clues to understand this ambiguity  are:  
 The  variables  describing `global properties' are mainly
decided by PQCD, not sensitive to the NPQCD properties, 
especially when averaged  for unbiased  events, as indicated 
by the hyperthesis local parton-hadron duality. 
Hence  jet shape variables,  inclusive momentum distributions 'should'  be similar for CS and CC, when averaged for 
UNBIASED events.  This indicate that we  must look into special events.
The baryon anti-baryon 
 rapidity  correlations which are  expected to be more sensitive to the NPQCD 
properties
however,  are also not able to 
expose the differences of the two colour connections.  So  other more sensitive observables need to be looked for. 
Another reason lies on the colour separate model we constructed . Only the singlet `glueball' cluster 
and its neighbouring 
2 strings of  the CS are different from  those  of CC, so the difference
is somehow  `localized' in a region of the phase space.  
This tells that the sensitive observables are more possibly defined in a 'window' of the phase space. 
The above arguments  
will be  much more clear after considering the following sections.

In this paper we 
selected two-jet like events, which will be described following.
First, taking  the advantage of generators, 
we select two-jet like events in the parton level,  i.e., by requiring 
the energy of 
all the gluons(including medium  state gluons)  smaller than a certain value.
  This part of events   
   shows that
CS and CC lead to hadron states with significant differences when measure 
a certain observable(section 2). 
Second, in section 3, we select 2-jet like events according to the transverse 
momenta of 
the hadrons, which can be done in experiment. This part of events shows that  
the colour separate singlet state can be  'probed': For these events, the CS
also  lead to significant differences from the CC,  
and comparing with data we can
extract the relative rate between the two kinds of colour connections
. All the numerical  analysize 
are performed with the
default JETSET7.4 (for CC) and the modified generator\cite{wangq}(for CS).
 The numerical results are for the $Z^0$ pole and the selected   two-jet like   events are of
the rate $\sim$ one per mill.

\section{Two-jet like events at parton level and a sensitive variable}

\begin{figure}
 \psfig{file=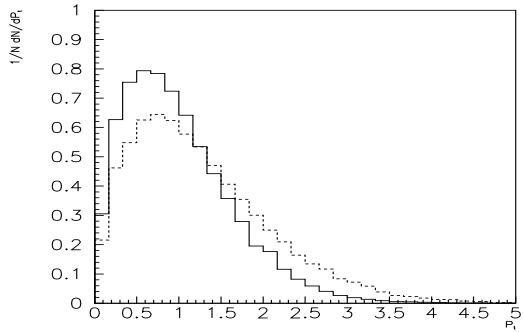,height=6cm,width=8cm}
\label{a4}
\vskip -0.8cm
\caption{The distribution of R  for two-jet like events selected at parton level.}
\vskip -0.6cm
\end{figure}

The parton level selected events and its properties are described in
\cite{lishiyuan}. Here  we define
\begin{equation}
R=\sum_i|p_{Ti}|,~~ \forall |y_i|<y_0
\end{equation}.
in which the $p_{Ti}$ and $y_i$ are the transverse momentum and rapidity of
the hadrons concerned.
This observable can 'accumulate' the various differences between  CS from CC. 
Its distribution for CS and CC is shown in Fig. \ref{a4}

The large differences  confirm us
that,  for the events we select, the CS and CC colour connections really
lead to large differences in hadron states. Hence we can say that they do  not
equal to each other from the point of view of hadronization/NPQCD. 
The differences  can not be smeared by tuning various parameters in the
generator\cite{lishiyuan}.

\section{Selecting two-jet like events at hadron level}

The discussions of the above section can be considered as a ideal experiment,
which confirm us that the different colour connections really leads to
different hadron states.
Now let's see if we can select such kind of events from experiments. 
By comparing the data and the predictions of different colour connections, 
we can extract the relative weight of them.

\begin{figure}
\psfig{file=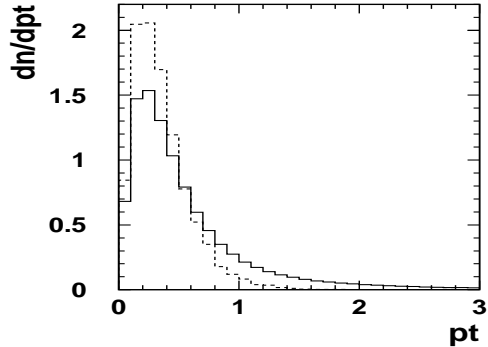, height=6cm, width=8cm}
\vskip -0.8cm
\caption{The $p_T$ distribution of final state hadrons for two-jet like
events which is selected at parton level(dashed line) and that for unbiased
events.}
\label{a10} 
\vskip -0.6cm 
\end{figure}

It is easy to understand that no direct correspondence to  the parton
level selected two-jet like events  from the point of view of experiment. The
crucial reason is that the set of events with the energies of all the gluons
in a event smaller than  a certain value have  no definite  meaning. It is
affected by, e.g., the cut-off of the parton cascade. At the same time,  
It is difficult to select these events with high purity by the jet shape
variables. However, we can see from Fig.\ref{a10} that the $p_T$ distribution
 of final state hadrons for these
 events is significant different from that of the unbiased events.
Especially in the region  $p_T<0.5GeV/c$, these
events can take a large part. 
Hence from now on we define our two-jet like events  as those events whose 
  final state
hadrons  all have the transverse momenta smaller than $0.5GeV/c$.
This can be measured in  experiments  and have definite physical meaning.

The most important thing is that,  the newly-defined events have very the
same properties as the parton level selected events\cite{lishiyuan}.

\begin{figure}
\psfig{file=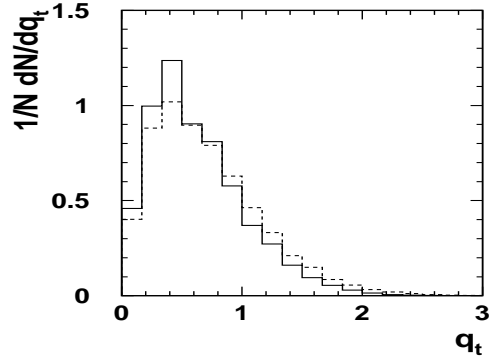, height=6cm, width=8cm}
\vskip -0.8cm
\caption{The R distribution for the hadron level defined two-jet
like events}
\label{a13}
\vskip -0.6cm
\end{figure}

Now we find that for the new-defined two-jet like events, there is large 
 differences for the R distribution. We can select these events, and compare
the data and theoretical prediction for variouis observables, the rate 
of different colour connections
can be extracted.  

\section{Discussions}

The conclusion of last section is very clear. One thing to emphasize is
that, The two-jet like events we defined in last section have never been
selected and compare with generator results until now. So when taking the comparison
to extract the rate of different colour connections, one should worry about
the validity of the generators. So what we have proposed here is also a very
interesting thing to do to test the generator.

The two-jet like events and the observable R 
looks similar as those in \cite{eden} at the first sight, But there are
crucial differences.  In that paper, for investigating hadronization,
the less gluons the better, however  we would like events with many soft gluons
so that  CS will have large probability. In that paper, they take a sum of
the vector $p_T$ to demonstrate different correlations among different
models, while we take the sum of the absolute value of the $p_T$ to
accumulate the differences.

\end{document}